\documentstyle[aps,prl,epsfig]{revtex}

\newcommand{\beq}{\begin{equation}}
\newcommand{\eeq}{\end{equation}}
\newcommand{\bea}{\begin{eqnarray}}
\newcommand{\eea}{\end{eqnarray}}

\begin{document}
\advance\textheight by 0.2in
\draft

\twocolumn[\hsize\textwidth\columnwidth\hsize\csname@twocolumnfalse%
\endcsname

\title{\bf Fluctuation and Dissipation in Liquid Crystal
Electroconvection} 
\author{Walter I. Goldburg$^1$, Yadin Y.
Goldschmidt$^1$, and Hamid Kellay$^{2}$}

\address{$^1$Department of Physics and Astronomy, University of
Pittsburgh, Pittsburgh, PA 15260\\
$^2$Centre de Physique Moleculaire Optique et Hertzienne\\ 
U. of Bordeaux,I, 351 Cours de la Liberation, 33405 Talence, France}
\date{\today}
\maketitle
\begin{abstract}
 In this experiment a steady state current is maintained through a
 liquid crystal thin film. When the applied voltage is increased
 through a threshold, a phase transition is observed into a convective
 state characterized by the chaotic motion of rolls. Above the threshold, 
 an increase in power
 consumption is observed that is manifested by an
 increase in the mean conductivity. A sharp increase in the ratio of
 the power fluctuations to the mean power dissipated is observed above
 the transition. This ratio is compared to the predictions of the
 fluctuation theorem of Gallavotti and Cohen using an effective
 temperature associated with the rolls' chaotic motion.
\end{abstract}
]
The fluctuation-dissipation theorem relates the spontaneous 
fluctuations in a system at temperature $T$ with its response to an 
external driving force.  An example might be a simple harmonic 
oscillator surrounded by air.  The air molecules exponentially damp 
its periodic motion and also cause the oscillating mass to fluctuate 
about its equilibrium position \cite{reichl80}.

The experiment described here is concerned, not with a system in 
thermal equilibrium but rather, one which is in a steady state: an 
electric potential $V$ applied across a weakly conducting liquid 
crystal (LC), generates convective motion within it when $V$
exceeds a critical value $V_{c}$ \cite{gleeson2001}. The LC is in good contact
with its surroundings, so that it's temperature is fixed at the ambient
value $T$, as it dissipates power $P$ at a mean rate 
$\left\langle P \right\rangle$.  Our 
interest centers on the dependence of $\left\langle P \right\rangle$
and its rms fluctuations 
$\sigma_{P}$ as a function of $V$
 or, equivalently, the control parameter 
 $\epsilon = ((V^2/V_{c}^{2})-1)$ \cite{dennin1996}.

Recently, Gallavotti and Cohen (GC) \cite{gallavotti95} have 
generalized the fluctuation-dissipation theorem (FDT) to include 
systems that are driven far from equilibrium, like the one studied 
here.  While their work motivated the present experiment, we are 
unable to verify its central prediction concerning the probability 
density function $\pi(P)$.  They showed that, under appropriate 
conditions, the system, assumed to be driven into a chaotic state,
experiences such large fluctuations in $P$ that sometimes $P$ can 
become negative, i.e. the driven, dissipative system momentarily  
can send energy back into the power supply that generates the chaotic 
fluctuations.  The theory makes a firm 
prediction about the ratio, $\pi (P)/\pi (-P)$. 

The GC theory is formulated in terms of the entropy production rate 
$s_{\tau }=\dot{S}_{\tau }$, which is related to the average power 
gained by the system during a time period $\tau $ by $s_{\tau 
}=P_{\tau }/kT_{ss}$.  Here the steady state temperature $kT_{ss}$ is 
equal to the mean  kinetic energy per particle.  
The authors show that if the chaotic motion in the system satisfies 
certain conditions, then
 $$\pi (s_{\tau })/\pi (-s_{\tau })=\exp(\tau s_{\tau }).$$
The theorem, 
which GC call the fluctuation theorem (FT), has been generalized by 
Kurchan \cite {kurchan98} to systems undergoing Langevin dynamics and by 
Lebowitz and Spohn to general Markov processes \cite{lebowitz99}.  The 
FT reduces to the FDT in the limit of vanishing driving force \cite{gallavotti96}.

For a macroscopic system like 
ours, it was not possible to achieve such large fluctuations in $P$
that render the power negative. Thus we were barred from directly measuring 
the ratio $\pi (P)/\pi (-P)$ predicted by the GC theory. 
What we did find were 
unexpectedly large values of the dimensionless standard deviation
$\sigma_{P}/\left\langle P \right\rangle$ when $V$ exceeded $V_{c}$.  
Our results become consistent with  predictions of the FT of GC, only if: 
(a) The temperature is taken to be the steady
state temperature $T_{ss}$ associated with the quasi-particles chaotic
motion and not the ambient temperature $T$ associated with the
microscopic thermal motion and (b) the correlation length of the velocity 
fluctuations is taken to be much larger than the typical size of the rolls.

The liquid crystal used here is methoxy benzylidene-butyl aniline 
(MBBA), a nematic that undergoes a transition out of the amorphous 
state, when driven by an ac voltage, which will be written as ${\cal 
V}(t) = \sqrt{2}V\sin(2\pi f t)$.  Typically $f$ is in 100 Hz 
range, and the measurements reported here were made at exactly this 
frequency.  The electrodes, across which this voltage is developed,
are thin, transparent layers of indium-tin oxide, enabling their 
visual observation.  When $V$ exceeds the critical value $V_{c}$= 6.0 
Volts in MBBA, the convective motion of the LC commences.  The 
resulting convective motion is analogous to Rayleigh-Benard convection, 
but with the applied voltage $V$ replacing the temperature difference
across the sample \cite{degennes1993}.

Most studies of electroconvection in LC's  center on the 
study of the patterns generated by the LC rather than the conductive 
behavior itself.  Observations through a microscope reveal that above 
$V_{c}$, the convective rolls form a stationary pattern. At a very 
slightly higher voltage, the dislocations develop and, with a further
voltage increase, commence to move.
At even higher values of $V/V_{c}$, the orientational fluctuations
take on a chaotic or turbulent appearance \cite{gleeson2001}
and the orientational domains fluctuate rapidly, though their motion is slow enough for the 
eye to follow. Thus this driven system, that appears to be in a chaotic steady state out of equilibrium,
is a candidate for checking the applicability of the FT.

In the present experiment,
the thicknesses $d$ of the MBBA samples were  37.5 and 50 $\mu$m. 
These samples 
were square with the  
lateral dimensions $L$ ranging from mm to cm. It was possible to make 
samples with $L$ less than 1 mm, but their 
resistance  was so high (hundreds of M$\Omega$) that 
measurements of the fluctuations in $P$ could not be reliably made.   

The experimental arrangement is extremely simple; the driving 
voltage ${\cal V}(t)$, obtained from a signal generator, is applied across 
the sample, of resistance $R$,  which is in series with a much smaller 
resistance $r$, which in our case was usually 100 K$\Omega$. The 
time-varying voltage $v_{r}$ is pre-amplified and then fed into a lock-in 
amplifier tuned to the driving frequency $f$. The lock-in output is
fed into dc amplifier (PAR 113) before it enters the 
A/D input of a computer, which 
records the detected signal. The current through the sample, 
$i=v_{r}/r$, and hence the power $P=iV$
fluctuates slowly compared 
to $1/f$, as we will see.  The full spectrum of the fluctuations can 
be captured by setting the lock-in time constant at $\tau =0.1$ s. 
 \cite{gleeson2001}.

Whereas this experiment is mainly concerned with fluctuations in $P$ about its 
mean value, this latter quantity itself 
exhibits an interesting dependence on the control 
voltage $V$.  Because $r<<R$, one might expect that
$\left\langle P \right\rangle=V^{2}/R$, so that a plot of 
$\sqrt{\left\langle P \right\rangle}$ $vs$ $V$ would yield a 
straight line of slope $1/R$, and indeed this is so.  But a more 
sensitive way of displaying the dependence of
${\left\langle P \right\rangle}$ on $V$ is to plot 
the sample conductance $G \equiv \left\langle P \right\rangle/V^{2}$ 
as a function of $V$.

\begin{figure}
\centerline{\epsfysize 4.5cm \epsfbox{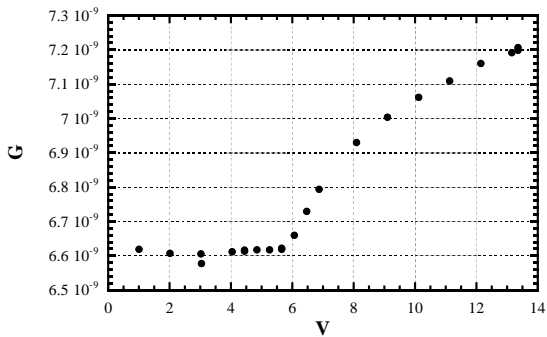}} \vspace{3mm}
\caption{Conductance 
$G=\left\langle P \right\rangle/V^{2}$, of MBBA  sample 
in inverse Ohms.  The control variable is the rms voltage, 
of frequency $f$ = 100 Hz.  The sample is a square 
of dimensions $L$ = 4 mm and  with $d$ = 50 $\mu$m.} 
\label{figure1}
\end{figure}
 
Figure 1 is a graph of this type.  For $V < V_{c}$ = 6 V, $G$ is 
independent of $V$, and the sample behaves like a resistor, with 
$R = 1/G = 1/[6.6 \times 10^{-9}]$ = 150 M$\Omega$. This resistor-like 
behavior was verified by replacing the sample with a metal-oxide film 
resistor having $R$ = 100 M$\Omega$. One sees from this figure that 
the convective motion generates increased energy dissipation in the 
sample when $V$ exceeds $V$ = 6 V. 

This appears to be the evidence
that a phase transition to a chaotic state enhances the 
entropy generation rate $s_{\tau}$ (see above).  The time 
interval $\tau$ in this experiment is the integration time of the 
lock-in, which is $\tau$ = 0.1 s. Note that the vertical scale in 
Fig. 1 spans a very narrow range in $G$.  It is not surprising then, 
that the transition at $V_{c}$ was not detected in a similar measurement of the 
conductivity as a function of $V$ in MBBA \cite{gleeson2001}.

\begin{figure}[tbp]
\centerline{\epsfysize 6cm \epsfbox{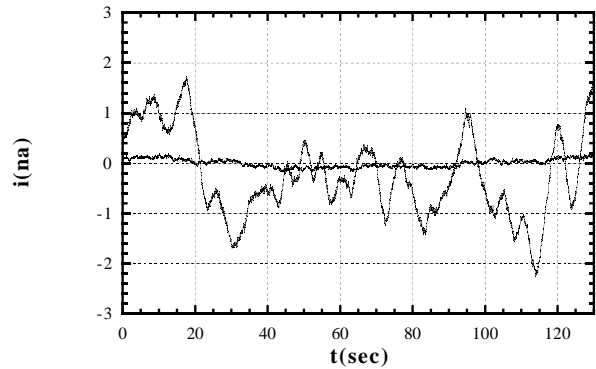}} \vspace{3mm}
\caption{Current through sample in nanoamps with 
its mean value subtracted out and a 
linear slow drift removed as well.  The strongly fluctuating
signal was recorded at  $V$ = 1.67 $V_{c}$.  The weak 
signal is at $V$ = 4 volts = 0.667 $V_{c}$.  
The lateral dimensions of this square 
sample are $L$=2.2 mm, with $d$= 37.5 $\mu$m.}
\label{figure2}
\end{figure}

 Turning now to the {\em fluctuations} in $P$ about its mean value,
 Fig. 2 shows  a time trace of $i(t)$  
 at two applied voltages: $V$ = 4.0 volts (weakly 
 fluctuating signal) and $V$ = 10.0 V (large, slow fluctuations).
 The sample thickness here is $d$ = 37.5 $\mu$m,  and its lateral dimensions are 
 2.3 mm and 2.15 mm. Over the two-minute interval of the measurements, 
 the mean current drifts slightly, an effect probably caused by a slight drift in room 
temperature. The sense of the drift varies in sign for different measurements.
To compensate for this drift, a term linear in $i$ is subtracted out. 

We have also measured the probability density function of the power 
dissipation $P$ at various values of $V$. There was no systematic 
departure from a Gaussian form for this function $\pi(P)$; its
kurtosis was quite close to the gaussian value of 3.  The 
accuracy of these measurements is limited by the slow drift in the 
lock-in output mentioned above.  It suffices then, to report the width 
$\sigma_{P}$ of this this function.

Figure 3 shows the non-dimensional standard deviation (STD) of the 
power fluctuations 
\[
\sigma _{P}/\left\langle P\right\rangle =\sqrt{\langle (P-\left\langle
P\right\rangle )^{2}\rangle/\left\langle P\right\rangle ^{2}} 
\]
as a function of $V$.  The sample is square with
$L$ = 4 mm and $d$ = 50 $\mu$m.

\begin{figure}[tbp]
\centerline{\epsfysize 5cm \epsfbox{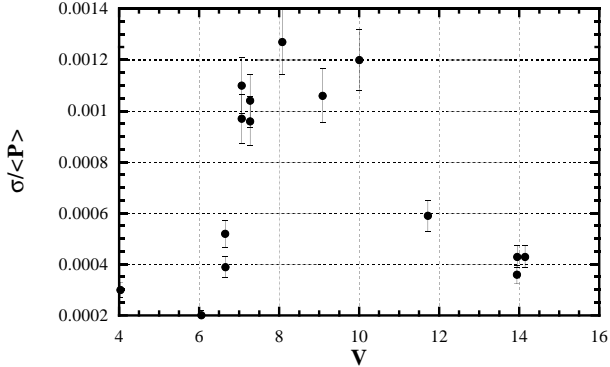}} \vspace{3mm}
\caption{Dimensionless standard deviation of the 
power fluctutations, $\sigma_{P}/\left\langle P 
\right\rangle$ $vs$ $V$.  For this sample $L$ = 4 mm and $d$= 50
$\mu$m.}
\label{figure3}   
\end{figure}

To interpret the power fluctuation data in Fig. 3, we assume that the
noise that appears at $V<V_c$ is uncorrelated with the much slower
fluctuations of interest above $V_c$ and, being of small magnitude, we
can neglect its contribution to $\sigma_{P}/\left\langle P 
\right\rangle$. We will also assume that above $V_c$ the bulk of the
power is dissipated as a result of the convective motion. 

In the voltage interval where $\sigma_{P}/\left\langle P \right\rangle$ 
is large, the fluctuation time is significantly longer than outside that interval.  
This observation was difficult to quantify, even though it is apparent from looking at
a record if $i(t)$.  For example, the fluctuation rate appears larger than 10 Hz 
(the inverse of the lockin time constant) at $V=$ 4 V, whereas it is smaller than 0.2 
Hz at $V$ = 8 V.  At 14 V the fluctuation rate has increased to roughly 1 Hz.

The FC theorem of Gallavotti and Cohen concerns the rate of entropy
production ${s}_{\tau }$ in thermostatted systems averaged over an
observation time $\tau $. The power dissipation can then be written as 
$P={s}_{\tau }T_{ss}$, where $T_{ss}$ is a temperature characterizing the
kinetic energy of the fluctuations in the chaotic or turbulent steady state.
Making certain assumptions about the time-reversal invariant chaotic motion
that is generated by the driving force, Cohen and Gallavotti obtained the
FT\ theorem as stated in the introduction. If we further assume that the
power fluctuations about the mean are gaussian, 
\[
\pi_\tau(P) \propto \exp[-(P-\langle P \rangle)^2/(2 \sigma_P^2))]
\]
as they appear to be, we find that
\[
\frac{\pi_\tau(P)}{\pi_\tau(-P)}=\exp(2 \langle P \rangle P
/\sigma_P^2).
\] 
Comparing this to GC result
\[
\frac{\pi_\tau(P)}{\pi_\tau(-P)}=\exp(\tau P/k T_{ss}),
\]
we find that the dimensionless STD is given by 
\begin{equation}
\sigma _{P}/\left\langle P\right\rangle =\sqrt{2k_{B}T_{ss}/\left\langle
P\right\rangle \ \tau }.
\label{STD}
\end{equation}

To interpret our measurements in the convective region, $V>V_{c}$= 6.0 V, we
will assume that the convective noise is uncorrelated with the noise 
that appears at, say, $V$ = 4 volts, which is well below $V_{c}$, and 
we subtract it out in a definition of the {\em excess} noise $E(V)$: 
\[
E(V)=[\sigma _{P}(V)-\sigma _{P}(V=4V)]/[\left\langle P 
\right\rangle-\left\langle P(V= 4V) \right\rangle] 
\]

To account for the measured maximum value 
of $\sigma _{P}/\left\langle P \right\rangle$ ($\simeq $ 10$
^{-3}$), we replace it with $E(V)$ and define a ``convective temperature'' $
T_{ss}$ by 
\begin{equation}
k_{B}T_{ss}=(1/2)Mv_{rms}^{2},
\label{mass}
\end{equation}
where $M$ is an appropriate mass of quasi-particles participating in
the chaotic motion, and
$v_{rms}$ is the characteristic velocity of the local convective motion
of the LC. We emphasize that $T_{ss}$ differs from the molecular 
temperature, which is very close to the ambient temperature ($\sim$ 
300 K), since the chaotic degrees of freedom are not in thermal 
equilibrium with the microscopic degrees of freedom.

By observing the trajectories of small particles convected by the
motion of the LC \cite{joets86,bodenshatz88}, one finds that, in order of
magnitude, $v_{rms}=v_{0} \epsilon$, where $v_{0}$ = 16 $\mu $m/s.
Our own admittedly crude measurements of the motion of dust particles give 
$v_{rms}\simeq $ 400 $\mu $m/s at $V$ = 9.6 V for the 4 mm sample used in
Fig. \ref{figure3}. 

The diameter $d$ of a roll is roughly the plate spacing, and in our 
samples their length was typically five times as large.  For the sample
 of Fig. 3 this gives $M$ in Eq. (\ref{mass}) $\simeq 10^{-9}$ Kg, since 
 the specific gravity of MBBA is close to unity.  Taking $v_{rms} \approx 100 $
 $\mu$m/s one obtains $2k_{B}T_{ss} = 10^{-7}$ J. Using $\tau$ = 0.1 s and the measured 
 value of $\left\langle P \right\rangle$ at $V$ = 10 V which is about
$7 \times 10^{-7} W$,  the rhs of Eq. (\ref{STD}) becomes approximately
$10^{-5}$, which is smaller by two orders of magnitude than the measured 
 $\sigma_{P}/\left\langle P \right\rangle$.  To accommodate for this difference,
we argue below that 
 the motion of the rolls may be correlated over scales much larger 
 than the size of an individual roll. 

The transition to the convective state is a continuous one as seen in Fig. 1;
it resembles a continuous magnetic phase transition in a system in thermodynamic 
equilibrium or a supercritical Hopf bifurcation \cite{rehberg}, 
The "ordered phase" occurs for $V>V_c$. 
Recently it has been claimed \cite{bramwell}
that a finite turbulent system is critical at all Reynolds numbers. It
thus resembles a two-dimensional XY model which is characterized by a diverging
correlation length throughout the ordered phase. This is because all
length scales between the size of the system, where the driving occurs,
to the microscopic scale, where dissipation occurs, are connected by 
the cascading process. Thus, we
would expect the system to exhibit a large correlation length (of the
order of the system size) above the onset of 
convection.  This may provide a possible explanation for the large 
magnitude of the fluctuations. If the rolls exhibit a collective correlated
motion, then the relevant mass could be as large as the total mass of 
the liquid crystal.  This will increase our estimate of $\Omega$ 
sufficiently to bring the measurement and the calculation  of $E(V)$ 
into approximate agreement.  The correlation length associated with 
the collective motion of the rolls should be distinguished from the much smaller length 
that is associated with the director fluctuations. This state of the LC in the present 
experiment is of course not turbulent, because the Reynolds number $Re$ associated with the roll size and 
$v_{rms}$ is much less than unity, but the motion is nevertheless chaotic.
Hence the present experiment is very different from that discussed in Ref. \cite{bramwell},
where the Reynolds number was large.  

To establish whether the numerator in Eq. (\ref{STD}) is indeed
proportional to the mass $M$ of entire sample, we measured 
$\sigma_{P}/\left\langle P \right\rangle$ and $\left\langle P
\right\rangle$ in two samples having different lateral dimensions, $L$
= 2 mm and 11 mm. Both measurements were made at the same voltage $V$
= 10.1 V, so that $v_{rms}$ presumably remains the same. According to the
 above model, $\sigma_{P}/\left\langle P \right\rangle \propto 
L/\sqrt{\left\langle P \right\rangle}$.  The ratio $R= 
(\sigma_{P}/\left\langle P \right\rangle)_{2mm}/(\sigma_{P}/\left\langle P
\right\rangle)_{11mm}=1/2$.  This is to be compared with the measured
ratio, $R$ = 2.  To resolve this inconsistency, we  must assume
that the correlation length of the power fluctuations saturates at
a scale of the order of a mm. With this assumption, the measured and
calculated values of $R$ come into approximate agreement.   

In summary, we have measured the steady-state power dissipation
$\left\langle P \right\rangle$ 
in a fluid system (thin layer of liquid crystal) 
in the vicinity of a phase transition from the quiescent to the 
convective state, where the fluid motion of the MBBA liquid crystal 
presumably becomes chaotic. To maintain 
this chaotic motion at voltages exceeding the critical value,
the system extracts energy from the power supply at an 
increasing rate, implying an increase in its mean conductivity $G(V)$.
 
The  dimensionless width of the fluctuations, 
$\sigma_{P}/\left\langle P \right\rangle$,
is much larger than that associated with
a resistor at ambient temperature $T$ but goes through a maximum 
near $V \simeq 2 V_{c}$. The magnitude of  $\sigma_{P}/\left\langle P \right\rangle$ 
near its maximum value can 
be approximately understood in terms of the fluctuation-dissipation 
theorem and the fluctuation theorem of Gallavotti and Cohen, provided
that one inserts into Eq. (\ref{STD}) a temperature $kT_{ss}$ that
is identified with the kinetic energy of the quasi-particles excited 
by the flow.  If the mass of these excitations is taken to be that of the 
convective rolls in the system, the dimensionless width of the power  
fluctuations is somewhat smaller than the measured value.  An even better
agreement is achieved if one assumes that the motion of the rolls is
correlated with a correlation length much larger than the size of the
rolls observed with a microscope. It thus appears that the mass of the
excitations in a measurement of the power fluctuations
is an appreciable fraction of the mass of the liquid crystal.

We are grateful for informative contacts with 
G. Gallavotti, E. G. C. Cohen, J-B Yoo, J. R. Cressman, X. L. Wu
and V. Horv\'ath.   
Without the help of J. T. Gleeson, this experiment could not have 
been carried out. This work is supported by the 
National Science Foundation under grants INT-9603361 and 
DMR-9701668 and by the U.S. Department of Energy (DOE), 
Grant No. DE-FG02-98ER45686.

Note added: After this work was completed we received a preprint from 
J. T. Gleeson et al. \cite{gleeson} showing a transition similar to that depicted in
Fig 1.

\end{document}